\RequirePackage[normalem]{ulem} 
\RequirePackage{color}\definecolor{RED}{rgb}{1,0,0}\definecolor{BLUE}{rgb}{0,0,1} 

\documentclass[letterpaper,twocolumn,10pt]{article}
\usepackage{usenix2019_v3}
\usepackage{hyperref}
\usepackage{longtable}
\usepackage{enumerate}
\usepackage{xparse}
\usepackage{subfigure}
\usepackage[]{algorithm2e}
\PassOptionsToPackage{hyphens}{url}\usepackage{hyperref}
\usepackage{booktabs} 
\usepackage{hyperref}
\usepackage{textcomp}
\usepackage{graphicx}
\usepackage[htt]{hyphenat}
\usepackage{mathtools}
\usepackage{amssymb}
\usepackage{nccmath}
\usepackage{longtable}
\usepackage{enumerate}
\usepackage{xparse}
\usepackage{color}
\usepackage[normalem]{ulem}
\usepackage{xspace}

\usepackage{enumitem}
\usepackage{url}
\usepackage{wrapfig}
\usepackage{lipsum}

\usepackage{caption}
\usepackage{etoolbox}

\usepackage{tikz}
\usepackage{amsmath}

\usepackage{filecontents}

\begin{document}

\date{}

  
\title{What breach? Measuring online awareness of security incidents by studying real-world browsing behavior}

\def\plainauthor{Author name(s) for PDF metadata. Don't forget to anonymize for submission!}

\author{
{\rm Sruti Bhagavatula}\\
Carnegie Mellon University
\and
{\rm Lujo Bauer}\\
Carnegie Mellon University
 \and
 {\rm Apu Kapadia}\\
Indiana University Bloomington
} 

\newcommand{\code}[1]{\texttt{#1}}
\newcommand\feature[1]{\vspace{.5\baselineskip}\noindent\emph{#1}}

\newcommand\featureshort[1]{\textsf{#1}\xspace}
\newcommand\datacomp{\featureshort{data\_compromised}}
\newcommand\precursortype{\featureshort{precursor\_type}}
\newcommand\homepage{\featureshort{precursor\_is\_homepage}}
\newcommand\firstoccurtype{\featureshort{1st\_occur\_type}}
\newcommand\firstoccursentiment{\featureshort{1st\_occur\_sentiment}}
\newcommand\breachnum{\featureshort{incident\_num}}

\newcommand\secref[1]{Sec.~\ref{#1}}
\newcommand\secrefs[2]{Sec.~\ref{#1}--\ref{#2}}
\newcommand\secrefssep[2]{Sec.~\ref{#1},~\ref{#2}}
\newcommand\appref[1]{App.~\ref{#1}}

\maketitle

\begin{abstract}
Awareness about security and privacy risks is important for developing good
security habits. Learning about real-world security incidents and data breaches can alert people to the ways in which their information is vulnerable online,
thus playing a significant role in encouraging safe security behavior.
This paper examines
1)~how often people read about security incidents online, 2)~of those
people, whether and to what extent they follow up with an action, e.g., by trying
to read more about the incident, and 3)~what influences the likelihood that they will
read about an incident and take some action. We study this by quantitatively examining real-world
internet-browsing data from 303 participants.

Our findings present a bleak view of awareness of security incidents. Only 16\% of 
participants visited any web pages related to six widely 
publicized large-scale security incidents; few read about one even when 
an incident was likely to have affected them (e.g., the Equifax breach 
almost universally affected people with Equifax 
credit reports). We further found that more severe incidents as well 
as articles that constructively spoke about the incident inspired
more action. We conclude with recommendations for specific future 
research and for enabling useful security incident information to reach more people.

\end{abstract}


\section{Introduction}
\label{sec:intro}

Security awareness is crucial for people to have the tools and
know-how for keeping their computers, accounts, and data safe~\cite{awareness-behavior2}.
With the rise of security incidents and data breaches, it is especially
important that people are aware of such incidents and their implications, whether or not they
were directly affected. Learning about past incidents informs people about the kinds of threats they may encounter
and as a result people are more likely to implement better security practices~\cite{awareness-behavior2}.
There have been almost 1,300
publicly reported breaches in 2018~\cite{bombshell-hacks-2017, 21st-cent-hacks, 2015-hacks, 2017DataBreaches, axel_axel_2018}.
High-profile cyber attacks in the past decade such as WannaCry, Heartbleed, Petya, and NotPetya 
have compromised over 300,000 systems worldwide~\cite{ransomware, vulnerabilities, wannacry_cleanup}.
The data compromised in breaches has ranged from passwords to personally identifiable 
information like names, 
email addresses, credit card numbers, and social security numbers. 
In addition to the need for affected people to become aware of incidents and take action,
it is also important for people to generally be aware of the extent and effects of security incidents as a stepping stone towards
preemptive  security practices~\cite{awareness-behavior}.

To this end, research about awareness of security incidents
has started to examine these issues
through surveys and
interviews, and has found that people learn about breaches from a variety of sources
and
that some breaches are more likely to be talked about than others~\cite{DBLP:conf/chi/DasLDH18}.
A survey
found that almost half the respondents had heard about a breach from a source
other than the breached company~\cite{ablon2016consumer}.
People's reported willingness to take action
was shown to be correlated with the source of information
about remediation steps~\cite{equifax-perceptions}, which people were more likely to take
if they perceived
a tangible security benefit~\cite{data-breaches}. Overall, these studies provide an important step
towards understanding how people learn about and react to breaches.
However, research thus far has
relied largely on participants' recollection of historical 
behavior or hypothetical situations, and so is constrained 
by common limitations of self-reported methodologies~\cite{ur2016users, habib2018away, wash2017can, 
rosenman2011measuring, fan2006exploratory, hanamsagar2018leveraging}.

In this paper, we take a significant step toward a more detailed
understanding of how people learn about and take action after incidents,
specifically through online browsing.
For a set of six national-scale security incidents of potentially varying relevance 
to people, we 
use \emph{longitudinal, real-world browsing data} to examine to what
extent people become aware of these incidents and the subsequent actions they take (e.g., by trying to
learn more about the incident or generally about security and privacy). Keeping in mind an underlying goal of improving
the spread of incident information through online media, 
we specifically study these problems in the context of \emph{online browsing},
without considering other channels through which which this information may be shared.
Our dataset was collected from the home computers of 303 participants between October 2014 and August 2018
and includes, among other types of data, all URLs visited and passwords used to log onto
online services from participants' home computers. We further conducted a follow-up survey of 109
participants asking about their device usage and knowledge
of security incidents to confirm that our longitudinal measurements
contain enough data to support our conclusions.
The results of this follow-up survey suggest 
that the browsing data in our dataset represent more than half of participants'
overall browsing (57\%). We discuss the results of this follow-up study 
in \secref{sec:validity} and
address the limitations of our data set in \secref{sec:limitations}.

We explore two main topics: 
First, we examine \emph{how often} people read about incidents
on the web 
and whether 
the likelihood of reading about incidents is associated with demographics,
browsing habits, or self-reported security behaviors.
Second, we seek to understand \emph{how people come to read} about incidents,
\emph{how they react} to finding out about them, and 
how the different ways
of finding out about incidents are related to how people take action.  For example, we
examine whether the type of web content (e.g., news vs.\ social
media)
on which we first observe people reading about incidents
affects whether
they take constructive action, such as further investigating an
incident or searching for instructional articles.

We found that
only 16\% of our 303 participants
visited an incident-related web page
 about any of six major security incidents between 2014 and 2017.
For example, only 15 of 59 likely Equifax credit-report
holders read about the breach online in our dataset.
Furthermore, these number remain alarmingly low even after accounting (through our confirmatory survey) for mobile browsing not captured in our data set.
Overall, we found that older and more tech-savvy participants were
more likely to read about security incidents on the
internet, as were participants
with higher self-reported proactive awareness
about their security~\cite{egelman2015scaling} 
and
participants who browse more technical or
technology-related web pages.

73\% of the participants whom we observed reading about an incident
subsequently visited additional web pages with information about the
incident or about security and privacy in general.  Reasonably, 
the higher the severity of the data compromised, the more likely
participants were to visit related web pages.
Participants' likelihood of taking action was higher if
the content through which they found out about the incident had
a positive sentiment;
no other property of the
breach-related content seemed to be associated with
taking action, even though our power analyses showed we had a sufficient sample
size to show medium-sized effects.

Overall, our results suggest remarkably low awareness of, or
inclination to follow up on, security incidents.
Our results also suggest
that information about security incidents does not reach online populations evenly
and that for those whom it reaches, the presentation of the information
can play a role in inducing action.
However, much work
remains both to help people become aware of security incidents and to help guide
them towards improved security and privacy hygiene.

We next survey related work (\secref{sec:related}) and describe our
dataset (\secref{sec:data_intro}).
We then describe the methodology and results for
the two main questions we investigate: \emph{how many}
and which people read about incidents (\secref{sec:exp_or_not}) and
\emph{how} people learn about incidents online and how this affects their actions (\secref{sec:learning_actions}).
We then briefly describe a follow-up study that substantiates our main results using self-reported data (\secref{sec:validity}).
Finally, we discuss the limitations and implications of our research
(Sections~\ref{sec:limitations} and~\ref{sec:discussion}).


\section{Related work}
\label{sec:related}

We survey three categories of related work:
security incidents and how people perceive them or react;
the dissemination of security and privacy information or advice and
its influence;
and methodologies for measuring security behavior.

\subsection{Awareness and perceptions of security incidents}
\label{sec:rel_work_breaches}
Much of the existing work about security incidents studies how people
interact with incidents such as data breaches, e.g., how people perceive data breaches and notifications
and the risks involved~\cite{data-breaches, equifax-perceptions},
what influences people to take action after a breach~\cite{equifax-perceptions, data-breaches}, 
and how people hear about data breaches~\cite{DBLP:conf/chi/DasLDH18, data-breaches, 
equifax-perceptions, ablon2016consumer}. 

Recent work found, for example, that people reported to be
more willing to take remediation actions if they perceived 
a tangible security benefit~\cite{data-breaches}.
Research that focused on
the Equifax breach
found that people reported that the source of advice about
steps to take after the breach played a role in their willingness to take 
action~\cite{equifax-perceptions}.
Researchers also measured customers' spending
habits before and after a breach announcement, and found that
spending was significantly reduced after an announcement
of a breach of a retailer's site~\cite{janakiraman2018effect}.
Other researchers found that only a small minority of survey participants
would stop business with a company after the company suffered a
breach~\cite{ablon2016consumer} and that almost half learned about
breaches from a source other than the affected company.
Previous work has also studied
people's general awareness of breaches and how breach information comes to people's attention
and found that social media accounted for almost a third of their participants'
information sources.~\cite{DBLP:conf/chi/DasLDH18}. More recent work 
surveyed people's reactions to 
notifications of password compromise.  
When advised or
required to change their passwords by the notification, less than a third of respondents reported any intention to change
their passwords~\cite{golla2018site}. Other recent work has shown that when a security incident occurs that involves
accounts on a major social network, people exhibit a variety of responses,
from doing nothing to actively seeking out information~\cite{redmiles2019should}.

Our work draws inspiration from previous work that examines how people
come across incident information and suggests that
the source of information is important for
taking action in response to a security incident.
We focus on incident information on web pages and base our analyses
on participants' real browsing behavior. 

\subsection{Information dissemination and influence}
\label{sec:rel_work_information_advice}
Related work also studied the
mechanisms and
sources from which people learn about security and
privacy,
often finding that social media and other web-based methods
are good channels for this task.
For example, prior work found that people use Twitter as a medium to complain or
share opinions regarding security incidents~\cite{DBLP:conf/soups/DunphyVTNMO15}.
Researchers have also found that people receive security and
privacy information through informal stories from friends and acquaintances
and that conversations about security
and privacy drive people to share with and advise others~\cite{DBLP:conf/soups/RaderWB12, 
  DBLP:conf/soups/DasKDH14}. Recent qualitative work 
found that older adults, in contrast to their younger counterparts, tend not to rely on
internet sources 
but rather on 
social resources such as advice from friends and family~\cite{nicholson2019if}.
Prior work has also studied themes in security and privacy
advice across three different sources---news articles, web pages with security advice, and informal stories from
family or peers---and found that each source presents information in a differently constructive way~\cite{rader2015identifying}.
Research also
found that the sources of
security and privacy advice
were important factors for people's digital security habits~\cite{DBLP:conf/sp/RedmilesMM16} (as also described in
\secref{sec:rel_work_breaches}), and
that the amount of advice that people reported receiving
was not distributed evenly among economic classes~\cite{DBLP:conf/ccs/RedmilesKM16}.
Researchers have also looked at the ways in which presenting people with security information
may help convince them to adopt good security practices.
One such work proposed interfaces for filesystems that show people how others
implement security~\cite{DBLP:conf/soups/DiGioiaD05}.
Similar work found that
showing people that their friends use security-enhancing features
on social networks increases the uptake of these features~\cite{DBLP:conf/ccs/DasKDH14}.

Our work is motivated by the findings that web-based media
are useful mechanisms for spreading computer security and privacy information.
With a long-term of goal of sharing such
information and advice 
more effectively,
our work specifically
aims to understand empirically how relevant information is consumed 
via web browsing.

\subsection{Measuring security behavior}
\label{sec:rel_work_sec_behavior}
Two main approaches have been used to measure security behavior: 
collecting self-reported data though surveys, interviews, or controlled experiments---e.g., people's behaviors
when exposed to internet attacks~\cite{DBLP:conf/ndss/OnarliogluYKB12}, password 
updating habits~\cite{habib2018user}, and willingness to take remediation 
measures after a breach~\cite{data-breaches, equifax-perceptions}---and
instrumenting users' computers to observe security behaviors---e.g., measuring password reuse~\cite{florencio2007large, DBLP:conf/ccs/PearmanTNHBCCEF17, 
DBLP:conf/ndss/DasBCBW14}
or
the presence of 
malware on people's computers~\cite{forget2016or}.
Since self-reported data can be prone to biases
and may not be representative of the reality of peoples' security 
and privacy~\cite{ur2016users, wash2017can, asgharpour2007mental, rosenman2011measuring, fan2006exploratory, habib2018away, hanamsagar2018leveraging}, we focus, unlike most prior research on security incidents,
on empirical measurement of actual behavior.

Previous work has measured
how often people update their computer systems~\cite{forget2016or, canfield2017replication, wash2017can}, 
what security settings they use
on their computers~\cite{forget2016or, canfield2017replication},
whether they are infected with malware~\cite{forget2016or,
  canfield2017replication}, and the presence of third-party
applications~\cite{wash2017can}.
Prior work has also measured how often people click on unsafe links~\cite{Sharif18Prediction, canfield2017replication}, their private-browsing~\cite{habib2018away} and
password-reuse habits~\cite{DBLP:conf/ccs/PearmanTNHBCCEF17, florencio2007large, wash2017can},
and whether they install security-enhancing extensions~\cite{wash2017can}.
We focus on security behaviors related to web usage,
as we are specifically studying the use of web-based media in spreading information.

Previous work that extracted security behaviors from real data has collected data in multiple ways.
One set of researchers partnered with an internet service provider that recorded all HTTP traffic of consenting
participants~\cite{Sharif18Prediction}.
Others asked study participants to install a tool that collected their system logs and information about the passwords
they entered on web pages~\cite{wash2017can}.

We leverage a data-collection infrastructure
called the Security Behavior Observatory (SBO) (described next, in \secref{sec:data_intro}), which 
captures detailed, real-world behavior of home computer users through
instrumenting their operating systems and web browsers~\cite{DBLP:conf/hotsos/ForgetKACCT14,
forget2016or}. Data gathered by the SBO has been used to study password reuse~\cite{DBLP:conf/ccs/PearmanTNHBCCEF17}, private
browsing~\cite{habib2018away}, and people's maintenance of their systems for 
security~\cite{forget2016or, canfield2017replication}.

\section{Data collection and dataset}
\label{sec:data_intro}


\paragraph{Data collection}
We obtained data collected as part of the Security Behavior
Observatory (SBO) project, a longitudinal study of the security
behaviors of Windows computer users~\cite{DBLP:conf/hotsos/ForgetKACCT14,
forget2016or, DBLP:conf/ccs/PearmanTNHBCCEF17}
from October 2014 to July 2019.
Data collected by the SBO
includes information about system configuration, system events,
operating system updates,
installed software, and 
browser-related data such
as browsing history, browser settings, and the presence of browser extensions.
To collect this information, participants' home computers were instrumented with
software that collects data
via system-level processes and
browser extensions.
Additional data related to passwords entered into web pages was collected starting January 2017
and only in the Google Chrome and Mozilla Firefox browsers. 

The SBO is approved by the ethics review board at Carnegie Mellon University.
Data collected by the SBO has been
used to study, for example, private-browsing habits~\cite{habib2018away},
people's ability to detect phishing attacks~\cite{canfield2017replication}
and password reuse habits~\cite{DBLP:conf/ccs/PearmanTNHBCCEF17, pwdbreach:conpro20}.

Our study is based on longitudinal data that was collected by the browser extensions.
In particular, we use the following two sets of data.

\vspace{.5\baselineskip}
\noindent
\emph{Browsing history:} The browsing data we analyze
	spans a subset of the whole SBO dataset from October 2014 to June 2018, encompasses
	$505$ participants, and covers participants' browsing using Google Chrome, Mozilla Firefox,
        and Internet Explorer. Participants
	enrolled in the SBO study on different dates and for different durations.
	We study a subset of $303$ participants who were active in the study
        at the time of when each of several security incidents was publicly announced (see \secref{sec:exp_or_not}).
        The average duration for
	which the 303 participants were enrolled was 505 days.
	This dataset includes information about every
        URL visited in the web browser, along with 
        page titles and timestamps. 
     
\vspace{.5\baselineskip}
\noindent
\emph{Password data:} This dataset
	spans from January 2017 to August 2018 and includes $233$ of the $303$ participants.
	The data includes information about every entry made into
        a password field in a web page, as determined 
	by a
        browser extension, including: a salted one-way hash of the password and
	the URL of the form in which the password was
        submitted.
        We filter this dataset to exclude passwords used during failed login attempts or 
        entered by a user other than the main computer user by replicating the 
        filtering process used by prior work that examined passwords collected through the SBO~\cite{pwdbreach:conpro20}.

\vspace{.5\baselineskip}
This browsing data is retrieved from participants' main 
computers. We assess the accuracy of our results in the context
of participants' overall browsing across multiple devices through a follow-up study
of 109 SBO participants. \secref{sec:validity} shows that this confirmatory study
appears to support our main findings.

We further
discuss the limitations of this dataset
in \secref{sec:limitations}.



\section{Who reads about security incidents}
\label{sec:exp_or_not}

Here we examine how many and which people visit
security-incident-related web pages, as well as
what factors
are associated with their likelihood of visiting such a page.

We focus on a selected set of security
incidents
(\secref{sec:part_breaches})
and model participants by their demographics and technical backgrounds,
self-reported security intentions, and internet browsing behavior
(\secref{sec:which_people}). We then report on
the relationship between these features and the likelihood that
participants visited pages related to security incidents (\secref{sec:exposed_not_results}).

\subsection{Methodology}

\subsubsection{Identifying who read about security incidents}
\label{sec:part_breaches}
We examine six security incidents that occurred between 2013 and 2017~\cite{bombshell-hacks-2017,
  21st-cent-hacks, 2015-hacks} such that we would expect most people
  to have read about \emph{at least one} incident. We selected these security incidents because they 
1)~were large-scale incidents (not affecting only a local population), 2)~spanned a variety of incident types
from personal financial data losses to company document leaks to 
cyber attacks
on home computers,
and 3)~were represented in our browsing history dataset.
These incidents are:

\begin{itemize}[topsep=.25\topsep,partopsep=.25\partopsep,itemsep=.25\itemsep,parsep=.25\parsep]
  \setlength\itemsep{.25\itemsep}
	\item \textbf{Equifax breach:} September 2017 breach of the credit reporting site
	that compromised the
	personal information of almost $150$
	million customers~\cite{equifax_breach}.
	\item \textbf{Uber hack:} Late 2016 breach that compromised
	the personal information of $57$ million Uber users~\cite{uber_breach}.
	\item \textbf{Ashley Madison breach:} Data breach on the
          affair-centric dating site
          in July 2015 and compromised around 33 million users' private information~\cite{ashley_madison_breach}.
	\item \textbf{Panama Papers:} 
	April 2016 breach of $11.5$ million files from the database 
	of the world's fourth largest offshore law firm, Mossack Fonseca~\cite{panama_papers_leak, panama_bbc}.
	\item \textbf{WannaCry:} Ransomware attack in May 2015 that initially affected 
	over $70,000$ computers in $99$ countries~\cite{WannaCry_breach, wannacry_cleanup}.
	\item \textbf{Yahoo!\ breaches:} Two breaches: one 
	in late 2014 affecting over $500$ million user accounts and another  in 2013
	affecting over $1$ billion user accounts~\cite{yahoo_breach_1, yahoo_breach_2}. It was
	later revealed that all user accounts were hacked~\cite{yahoo_date_2}.
\end{itemize}
        
Each incident we study may be relevant to users in different ways, e.g., they could have been affected by it,
they could be users of the compromised service and may want to be more cautious in the future, or they could learn about security and privacy dangers
in a broader context based on the incident. For example, although Panama Papers may not be directly relevant to most users,
we included it because awareness about it could indirectly
encourage users to be cautious about the safety of their own private records (e.g., medical records) and maybe be selective
in trusting institutions with their data.
        
To study who reads about these incidents, we focus on the participants who were active
in the study before the incident became public and for three months after. 

For each incident,
we identified participants who
visited an incident-related page (henceforth we may call this \emph{reading about a breach}).
This web-page visit could be the first exposure to the incident or
an attempt to learn more about the incident online after hearing about it elsewhere.
Since we seek to study
how often people actually signal the intent to learn more about an incident
rather than simply ``hearing about it,'' we do not consider a participant to have 
read about an incident if it was mentioned on 
some web page, e.g., on social media, but the participant did not click on the article. 

To determine whether a
participant read about an incident, we performed a keyword search over
the URLs and titles of all the pages in their browsing
history.
For each incident we manually selected a set of keywords
that we believed would identify web pages that focus on that
incident. For example, we searched for various combinations of 
``Yahoo''
and one of the following: ``compromise'', ``attack'', ``breach'', ``hack''. To confirm
that our keyword lists were inclusive enough, we also performed multiple
Google searches using a variety of search terms to find web pages about
the incidents and then confirmed that each of the top 100 Google search results about each
incident would be identified by our keyword lists. We then manually verified that each page visit that matched
a keyword actually corresponded to
a page about the incident. For example, a page on the domain of \code{yahoo.com} with the path containing
the word ``hack'', referring to a page about life hacks would not be considered
an incident-related page.

\paragraph{Equifax and Yahoo!\ users}
\label{sec:equifax_yahoo_meth}
To provide further context for our observations of how many participants
read about an incident, we observed for people who were likely
to have been \emph{affected} by an incident, how many of them read about the incident
as part of our analysis.
Equifax and Yahoo!\ are the two breaches for which we
were able to relatively accurately
estimate how many participants were \emph{actually affected} by examining
whether they logged in to certain web sites. In both cases, the number of affected people
was all or almost all users or consumers~\cite{yahoo_date_2, fortune-equifax, equifax_breach}.

We determined people
who were likely to have had an Equifax credit report by those 
who entered passwords on credit-card reporting sites
that reported on Equifax credit ratings. 
To determine which participants were likely to have a credit report with Equifax (and were hence likely to be affected
by the breach~\cite{equifax_breach}), we searched for participants who had entered a password on
\code{annualcreditreport.com}, \code{creditkarma.com}, \code{equifax.com}, 
or \code{identityguard.com} before September 7, 2017, which is when the breach
became public. 
We picked these domains because they were
on a list of six popular credit-report sites
reporting Equifax scores~\cite{catalano_2018}
and appeared as domains on which participants entered passwords
in our dataset.
While most Americans were likely to have been affected~\cite{fortune-equifax, equifax_breach} regardless of
whether they had an account with a credit-reporting site, for this analysis, 
we considered this set of participants that were \emph{very likely} to have been affected according to the above criteria.

Similarly, to determine which participants had a Yahoo!\ account, 
we searched for participants who had entered a password on the
\code{yahoo.com} domain before February 15, 2017---when the
breach had first become public.
We repeated this search for participants who had a Yahoo!\ 
account before the second 2017 breach announcement, October 3, 2017.

Since we had access to passwords only for
Chrome and Firefox users, these are the only participants that we can
judge were affected by the Yahoo!\ or the Equifax breach.

\subsubsection{Studying which people read about incidents}
\label{sec:which_people}
After determining which participants read about at least one of the
incidents,
we study what characteristics of participants are correlated
with participants visiting pages related to the security incidents. We model
participants and their behavior using three distinct feature sets
and
then perform a logistic regression for each feature set, where the
outcome variable in each regression is a binary variable indicating
whether a participant read about an incident.

\paragraph{Feature set 1: Demographic characteristics}
Based on findings from previous work showing that demographics
were correlated with
how people share security and privacy news and their
comfort with uses of breached data~\cite{DBLP:conf/chi/DasLDH18, data-breaches},
we hypothesized that certain demographics would
also be correlated with whether they read about a security incident.
Therefore, the first feature set contains demographic information
about the participant:
age, gender, income, highest education level, whether the participant is a
student, whether the participant's primary profession involves
programming, and whether the participant knows at least one
programming language.

\paragraph{Feature set 2: Self-reported security intentions}
Prior work found self-reported security intentions (as
measured by the SeBIS scale)
to be correlated with how people heard about and shared security and
privacy news~\cite{DBLP:conf/chi/DasLDH18}. Hence, our second feature set 
comprises the
four continuous feature values of the SeBIS
scale~\cite{egelman2015scaling}, which participants optionally filled
out upon enrollment in the SBO.
The four values represent the extent to which participants a)~secure their
devices,
b)~generate strong and 
varied passwords across accounts, c)~demonstrate proactive awareness of security issues
or safety of websites and links, and d)~update the software on their computers.

\paragraph{Feature set 3: Participants' observed internet behavior}
We hypothesized that the types of web pages people browse would be correlated with
people's likelihood to encounter information about a security incident. For example,
we hypothesized that people who browse more technology-related news articles may
be more likely to come across web pages about security incidents.
To test these hypotheses, we examined two types of internet behaviors:
the kinds of topics of web pages that participants typically
visited and the amount of their web browsing that involved visiting web
pages on technical topics. We
describe each of these next.

\feature{Characterization of browsing behavior}: We used a topic-modeling algorithm to
generate a set of topics that categorize
participants' browsing. To generate the set of topics (which was the
same for all participants), we looked up the category of the domain of
every web-page visit in Alexa Web Information Services, resulting in a
multiset (i.e., bag) of words. We performed topic modeling using the
Non-negative Matrix Factorization (NMF) algorithm~\cite{lee1999learning} on that multiset, 
which identified two topics as
the most coherent
for modeling participants'
browsing (by determining the number of topics
with the highest intra-topic cosine similarities~\cite{stevens2012exploring}). 
One topic appeared to correspond to browsing that was
professional or work-related; the other topic related to browsing that was
leisure- or entertainment-related. (See \appref{app:browsing_behavior} for more details.)
The output of NMF also includes, for each topic and participant, a
value that describes how much of that
participant's web browsing matches the topic. Hence, each
participant's browsing behavior is characterized with two features
(corresponding to two topics).

\feature{Amount of technical content browsed}:
We further characterize people's browsing according
to how many of the web pages they visit are technical or technology-related.
We again use the NMF algorithm to build a topic model
with two distinct topics: one topic covering technical or technology-related
content and the other comprising all other types of content. This topic model
is computed over the content of web pages visited by the participants.
We consider a web page
to be technical or technology-related if the AWIS category
for the domain of the web page contains the word ``technical'' or
``technology''; otherwise we consider it not technical or
technology-related. 
For a sample of each participant's browsing history,
we download the content of the web pages in the sample using the newspaper library~\cite{newspaper}
and train a topic modeling algorithm to learn two topics based on two documents: the content
of downloaded web pages with a technical AWIS category and the downloaded content
of all other web-pages with a non-technical AWIS category. We then apply this topic model
trained for two topics on the multiset of tokens of downloaded content for each participant's browsing sample.
Similarly to when characterizing browsing behavior, the NMF algorithm outputs, 
for each topic and participant, a weight for the topic within the sample
of the participant's
browsing history. We characterize the
amount of technical content a participant browses by the weight corresponding to
technical content. (See \appref{app:browsing_behavior_technical} for more details.)

\subsection{Results}
\label{sec:exposed_not_results}
Our filtered set of participants includes $303$ participants who were
active in the study around the time of at least one incident announcement. Their ages
ranged from 20 to 83 years, with a median of 29 and a mean of 36. 59\% of
participants identified as female, 41\% as male, and
one did not provide their gender.
63\% of participants completed a Bachelors degree or higher.
48\% of participants were students.
30\% of participants had an income above 50,000 USD, 50\% below 50,000 USD, and the rest
did not provide their income.

We were surprised to discover that only 48 of the 303 (16\%)
visited a web page that discussed \emph{any} of the security incidents that we
studied.%
\footnote[1]{While we do not expect all participants to be interested in every incident,
it was likely that the majority of participants 
would find at least \emph{one} incident relevant to them in some way.}
Henceforth, we may say that these participants ``read about
the breach,'' even though we cannot confirm they 
understood the content of the pages they visited.
In three additional instances
participants searched for incident-related keywords
but did not visit any of the search results or any other incident-related page.
Table~\ref{tab:users_breaches} shows how many participants
visited a page about each security incident.

This fraction of participants is computed based on
browsing history from participants' home computers; our confirmatory
study suggests this accounts for the majority of
participants' web browsing (see \secref{sec:validity}).

We also examined a subset of participants that we hypothesized
were particularly likely to have been affected by the Equifax or
Yahoo!\ breaches (see \secref{sec:equifax_yahoo_meth}).
Of the 59 participants who entered passwords on credit-card reporting sites that reported on Equifax credit ratings,
15 (25\%) read about the Equifax
breach.  In contrast, of the 48 participants who entered a
Yahoo!\ password in the relevant time periods---and hence were
definitely affected by the breach~\cite{yahoo_date_2}---only one (2\%) read about the breach.
We substantiate these numbers through a follow-up study related to 
people reading about the Yahoo!\ and Equifax breaches online (\secref{sec:validity}).

We analyzed the relationship between the binary outcome of whether a participant read
about any of the incidents and each of the three feature sets described 
in \secref{sec:part_breaches} by computing three logistic regression 
models. When interpreting results, we used a significance level of $0.05$.

First, we computed a model exploring the effect of demographic
characteristics over the 303 participants (Table~\ref{tab:exposed_not_lr_demo}). 
We found that
participants' ages and whether they know a programming language were{\parfillskip0pt\par}

\begin{table}[]
\centering
\begin{tabular}{|l|l|}
\hline
\emph{Incident} & \emph{\# participants} \\ \hline
Equifax & 26 \\
Yahoo!\ & 6 \\
Uber & 4 \\
Ashley Madison & 6 \\
WannaCry & 14 \\
Panama Papers & 10 \\ \hline
\end{tabular}
\caption{Number of participants who read about each security incident;
  some
  read about multiple incidents.}
  \label{tab:users_breaches}
\end{table}

\noindent significant factors.
Specifically, older and more technology-savvy participants were more likely
to read about incidents. The effect of age was only marginal---the odds
of reading about an incident increased by 
 $1.003\times$ ($p=0.02$) for each additional year of age---but
 the odds of reading about an incident
increased by $1.149\times$ ($p = 0.002$)
if a participant knew a programming language.

Our second model examines the relationship between whether
participants read about an incident and their self-reported SeBIS scale values (Table~\ref{tab:exposed_not_lr_sebis}).
This model was computed over 247 participants who provided SeBIS data
to the SBO at the time of enrollment.
Only one of the values
of the four SeBIS score categories was statistically significant, which
we model by its Z-score for easier interpretation;
the odds of reading about an incident were increased by a factor
of $1.078$ ($p = 0.05$) for each standard deviation increase
in the SeBIS proactive awareness score
of a participant (a value in $[0, 1]$). 

Our third model examines the relationship between whether 
participants read about an incident and
their internet browsing behavior (i.e., browsing topics and amount of technical
browsing; see \secref{sec:which_people}).
This model was computed over 302 participants
who had enough browsing data from which a sample sufficient for 
computing the technical browsing descriptor could be drawn
(see \appref{app:browsing_behavior_technical}).
Of the factors examined by this model, only the amount of
technical or technology-related browsing 
was a significant factor (again modeled by its Z-score).
The
odds of reading about an incident are increased
by a factor of $0.026$ ($p < 0.001$) for every standard deviation increase in
the technical browsing score. Table~\ref{tab:exposed_not_lr_internet_behavior} contains the results of this logistic regression model.

\begin{table}[tp]
\centering
\resizebox{\columnwidth}{!}{
\begin{tabular}{|l|r|r|r|r|r|r|r|}
\hline
\textit{\textbf{}} & \textit{\textbf{baseline}} & \textit{\textbf{coef.}} & \textit{\textbf{exp(coef.)}} & \textit{\textbf{std.err.}} & \textit{\textbf{t}} & \textit{\textbf{p}} \\ \hline
(Intercept) &  & -0.070 & 0.933 &  0.074 &  -0.953 & 0.341 \\ 
\textbf{age} & \textbf{} & \textbf{0.003} & \textbf{1.003} & \textbf{0.001} & \textbf{2.311} & \textbf{0.021} \\ 
gender: male & female & 0.025 & 1.026 &   0.035 & 0.715 & 0.475 \\
Education: $\geq$ ugrad & \textless ugrad & 0.026 & 1.027 &  0.035 &  0.748  & 0.455 \\
Income: \textgreater \$25k & \textless \$25k & 0.002 & 1.002 & 0.040  & 0.053  & 0.957 \\
Income: declined to answer & \textless \$25k & -0.031 & 0.969 & 0.056 &  -0.562 & 0.574 \\
\textbf{knows\_prog\_lang: yes} & \textbf{no} & \textbf{0.139} & \textbf{1.149} & \textbf{0.045} & \textbf{3.084} & \textbf{0.002} \\ 
is\_programmer: yes & no & -0.028 & 0.972 & 0.049 & -0.583 &  0.560 \\
is\_student: yes & no & 0.050 & 1.052 & 0.047 &  1.078 & 0.282 \\ \hline
\end{tabular}}
\vspace*{1mm}
\caption{Logistic regression model describing the relationship between whether a participant
learned about a breach and characteristics of the participant including their demographics.
``ugrad'' denotes that the participant indicated achieving a
Bachelor's degree.}
\label{tab:exposed_not_lr_demo}
\end{table}

\begin{table}[t]
\centering
\resizebox{\columnwidth}{!}{
\begin{tabular}{|l|r|r|r|r|r|r|r|}
\hline
\textit{\textbf{}} & \textit{\textbf{coef.}} & \textit{\textbf{exp(coef.)}} & \textit{\textbf{std.err.}} & \textit{\textbf{t}} & \textit{\textbf{p}} \\ \hline
(Intercept) & -0.180 & 0.835 &  0.178 & -1.013 &  0.312 \\ 
Device Securement & 0.000 & 1.000 & 0.018 &  0.009  & 0.993 \\ 
Password Generalization & 0.020 & 1.021 &   0.039 &  0.535  &  0.593 \\
\textbf{Proactive Awareness} &  \textbf{0.075} &  \textbf{1.078} &  \textbf{0.038}  & \textbf{1.990}  & \textbf{0.047} \\
Updating & 0.002&1.023 & 0.023  & 0.969 &  0.333 \\ \hline
\end{tabular}}
\vspace*{1mm}
\caption{Logistic regression model describing the relationship between whether a participant
learned about a breach and the SeBIS scale values they provided.}
\label{tab:exposed_not_lr_sebis}
\end{table}

\begin{table}[t]
\centering
\resizebox{\columnwidth}{!}{%
\begin{tabular}{|l|r|r|r|r|r|r|r|}
\hline
\textit{\textbf{}} & \textit{\textbf{coef.}} & \textit{\textbf{exp(coef.)}} & \textit{\textbf{std.err.}} & \textit{\textbf{t}} & \textit{\textbf{p}} \\ \hline
(Intercept) & 0.060 & 1.062 &  0.073 & 0.823 & 0.411\\ 
Browsing: leisure & 0.401  &  1.493 & 0.354 &  1.132  & 0.258 \\ 
Browsing: professional & 0.635  &  1.887 &   0.420 &  1.510 & 0.132 \\
\textbf{Z(Browsing: technical)} &  \textbf{0.066} &  \textbf{1.068} &  \textbf{0.024}  & \textbf{2.690}  & \textbf{0.008} \\ \hline
\end{tabular}}
\vspace*{1mm}
\caption{Logistic regression model describing the relationship between whether a participant
learned about a breach and characteristics of their internet browsing behavior. }
\label{tab:exposed_not_lr_internet_behavior}
\end{table}

\subsection{Summary of findings} {Overall, participants who were older, exhibited a higher proactive awareness about computer security, and who were more technology-inclined were more likely
to come across information about security incidents online. This indicates an imbalance in the dissemination of important security and privacy information to
different demographics and user populations.}


\section{How people learn about incidents and take action}
\label{sec:learning_actions}

We now study how the 48 participants who read about security incidents
came
to visit incident-related web pages
and what behavior
they exhibit in response.
We first explain how we characterize reading about (discovery) and taking action after
an incident (\secref{sec:learning_actions_all_meth}); we then
examine how the characteristics of discovery or of
the incident relate to participants' reactions
(\secref{sec:learning_actions_all_results}).

\subsection{Methodology}
\label{sec:learning_actions_all_meth}

We define features that characterize the process of discovery
of web pages about incidents (\secref{sec:learning_meth}) and we characterize
participants' actions after discovery (\secref{sec:actions_meth}).

\subsubsection{Learning about incidents}
\label{sec:learning_meth}
We examine the \emph{browsing trajectories}---sequences of page visits
that surround the visit of an incident-related web page---of each
participant for each incident.
We then measure
the
characteristics of the page visits that were part of the trajectory
before the visit to a incident-related web page, and, separately, the
characteristics of page visits after the first visit to the
incident-related web page.
We analyze how the actions people
take---as observed by examining the part of the trajectory after
first visiting the incident-related web page---are related to characteristics of the incident, characteristics
of the browsing path up to reading about the incident, the
participants' demographics, and internet browsing behavior.

We constructed the browsing trajectories
by the following steps:
We first identify each participant's visits (if any) to web pages related to any of the incidents
as described in \secref{sec:part_breaches}.
For each \emph{first occurrence}---the first visit to any incident-related page
  about a specific incident---we define a trajectory to be composed
  of the 20 page visits immediately preceding this first visit to a
  incident-related page, the actual first visit to the
  incident-related page, and the 20 page visits that immediately followed.

In this manner, we construct one trajectory per incident for each participant
who visited any page about that incident.

To study how people read about incidents through browsing and their subsequent actions, we define
and analytically examine several features:

\feature{Precursor web page type (\precursortype)}
This feature describes the type of web page on which the participant
clicked on a link that took them to the first occurrence of a page 
about the incident. This is commonly called the ``referrer'' page; we
call it the precursor page because we identify these pages manually instead
of via referrer headers, which are often not available.
To create this feature, we manually categorized all the
precursor pages as follows:

\begin{itemize}[topsep=.1\topsep,itemsep=.2\itemsep]
	\item \textbf{Social media:} A social media site
          (e.g., Facebook) home page or a social media
  	  page related to the topic.
	\item \textbf{Message boards:} A message forum
          such as 4chan message boards, reddit, or a Charles Schwab online forum.
	\item \textbf{News page:} The page was on a news website.
	\item \textbf{Purposeful:} Search
          engine results about the incident.
	\item \textbf{General browsing:} The page did not fall into
          one of the above categories, but contained a link to the
          first-occurrence page. E.g., a stackexchange page
          with a sidebar link to an incident-related page.
      \item \textbf{Unknown:}
          No pages in the trajectory
          preceding the first occurrence of an incident-related page
          appeared to have a link to that page. This could happen if
          the participant entered the URL of the incident-related page
          manually or clicked on a link
          in an external tool.
\end{itemize}

In some cases there was an automatic URL redirect between
the precursor page and the first occurrence of a page about the
incident. Our analysis
correctly identifies 
the page that generated the redirect link as the
precursor page.

\feature{Whether the precursor page was a home page
  (\homepage)}
This feature captures whether the precursor page was the home page of
a domain or whether the participant had to have browsed more deeply
into a website before encountering the precursor page. We examine
this feature to determine whether the link to the first-occurrence page
was easily visible to anyone (i.e., on a home page) or 
would be seen only by some visitors to that site (i.e., those
who had navigated to a specific section).

\feature{First-occurrence page type
  (\firstoccurtype)}
This feature categorizes the
first-occurrence
page according to whether it is specifically about the incident,
and, if so, whether it is descriptive or prescriptive.
We used the 
newspaper library~\cite{newspaper} to extract the main content of each
page; we then manually examined the content, and developed the following
three categories, using which we then classified each page:
(1) general information about this specific incident, e.g.,
          what caused it;
(2) advice about this specific incident, e.g., what to do in
          response or how to find out if one is affected;
(3) not specifically about this incident, but 
          mentions the incident, e.g., a political article that mentions
          the incident.

\feature{First-occurrence page sentiment (\firstoccursentiment)}
Inspired by research on how the sentiment of social media posts
influences the sentiments of the poster's followers~\cite{bae2011sentiment}, 
we hypothesized that
people's reactions to web pages about incidents might be related to the
sentiment of the pages, in particular, that
positive
sentiment might correlate with more constructive action.

We computed the sentiment for the main content of each
first-occurrence page (collected as described above) using the
NLTK Vader library~\cite{gilbert2014vader}. 
This feature has values in the range $[-1, 1]$; lower values indicate
more negative sentiment, higher more positive sentiment, and 0 
neutral sentiment.

\feature{Incidents previously read about (\breachnum)}
We hypothesized that people react to incidents differently depending on
how many incidents they have come across through web browsing. Hence, for each incident
that a participant read about, we counted how many other incidents of the six incidents they
had read about previously (i.e., the number of trajectories
previously constructed for this participant for other incidents) and exposed this as a feature in our analyses. 

\feature{Type of data compromised (\datacomp)}
This feature represents the type of data compromised in the incident. We broadly group the data types and incidents as follows.

\begin{itemize}[topsep=.25\topsep,itemsep=.25\itemsep,parsep=.25\parsep]
	\itemsep0em
	\item {\bf PII:} names, phone numbers, partial credit card numbers, email or physical addresses (Ashley Madison, Uber);
	\item {\bf PII++:} 
	as above, and credit card information or
          social security numbers (Equifax)
	\item {\bf Passwords} (Yahoo!)
	\item {\bf Miscellaneous} (WannaCry, Panama Papers)
        \end{itemize}

Although passwords for many Ashley Madison users were eventually
cracked and leaked, this did not happen until months after the
original leak became public~\cite{ashley_madison_breach}; hence, we categorize the
Ashley Madison breach as only including PII and not passwords.        

\subsubsection{Actions after reading about incidents}
\label{sec:actions_meth}

We first describe how we determined the actions participants
took
after they read about an incident, which we use as
the outcome variables
in statistical analyses (\secref{sec:stat_results}).

We manually examined the 20 page visits in each trajectory immediately following
the first occurrence, as well as any visits to incident-related web pages \emph{after} the first
visit. We call one of these page visits an \emph{action} taken in
response to reading about the incident if it falls into at least one of the
following categories:

\begin{itemize}[topsep=.1\itemsep,itemsep=.1\itemsep,parsep=.5\parsep]
	\itemsep0em
	\item \textbf{Educating themselves about the incident:}
          e.g., reading additional articles about how the incident occurred, who is
          responsible, or implications of the incident.
	\item \textbf{Educating themselves about general security:} e.g., reading articles about how to secure
	their network or whether using personal emails for work is safe.
	\item \textbf{Taking action to make themselves more secure:} e.g., attempting to freeze their credit
	reports, visiting a website to download patches after a cyber attack,
	or reading ``what you need to do'' articles.
\end{itemize}

We then counted the number of actions after reading about an incident.
If a participant's browsing history included multiple page visits within 30 seconds 
whose titles and URLs were identical,
we counted them as one page visit.
We use this raw count of actions as the outcome variable
in our analyses (see \secref{sec:stat_results}).
For example, if a participant 
visited two more pages that discussed the breach as well as
one page with a ``what you need to do'' article, this would count as
having taken three actions after reading about the breach.

So as to treat incidents uniformly, we do not consider actions tailored to any specific incident (e.g., changing passwords
after a password breach).

\subsection{Results}
\label{sec:learning_actions_all_results}

In \secref{sec:descriptive_results}, we describe
how participants came to read about incidents and their actions after
reading.
In \secref{sec:stat_results}, we report on analyses that model the amount of action participants
take in relation to how they come across incident-related pages, the type of incident,
characteristics of participants, and their web-browsing behaviors.

\subsubsection{Descriptive results}
\label{sec:descriptive_results}
Using the methodology described
in \secref{sec:learning_meth},
we identified
66 distinct trajectories across the 48 participants (out of 303) who visited an
incident-related web page. 
About twice as many trajectories described 
a participant reading about a PII++ breach (26) than a PII breach (10), and about four times
as many described a participant reading about a PII++ breach 
than about a password breach (6). The remaining trajectories
(24) were about WannaCry and Panama Papers.

The types of
web pages that led participants to visit the first
incident-related page (\precursortype) were relatively evenly
distributed across social media (11), message boards (9), news pages
(13), searching for the breach (9), and
general browsing (10).
For 14 trajectories we could not
identify the page that led participants to their first visit of an
incident-related page.
For the 52 
trajectories for which we could determine the precursor
page, approximately half (24) the precursor pages were home
pages (\homepage); the others (28) were pages deeper in a website.

When we categorized the first incident-related page visits according to
their content (\firstoccurtype), we found that 10 were advice articles, 35
were pages with
general information about the incident, and 21 had content 
related to the incident (e.g., a story about a woman's identity stolen 15 times after the Equifax breach) 
without specific information about the incident.
The sentiment of the content of the first-visited incident-related pages (\firstoccursentiment)
was slightly positively skewed, with
a mean of $0.13$ and a standard deviation
of $0.73$. (Values above 0 indicate positive
sentiment; below 0 negative sentiment; and 0 is neutral.)~\cite{gilbert2014vader}).

Most participants (73\%) visited pages about only one of the six breaches
we focused on. Only 10 visited pages
about two breaches, one about three breaches, and two participants
visited pages about four distinct breaches.

\begin{figure}[tbp]
\centering
\includegraphics[width=\linewidth]{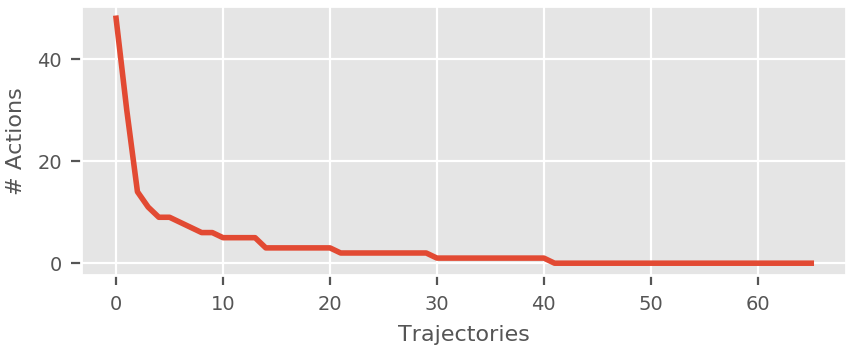}
\caption{Number of actions taken per trajectory. \label{fig:num_actions_trajectories}}
\vspace{-4mm}
\end{figure}

Most participants (73\%) who visited a breach-related web page afterward took
at least one action (i.e., visited another page about the breach, a
page about security in general, or a page describing how to
react to a breach). The mean number of actions taken across the $66$ trajectories was $3$ with a standard
deviation of $7.19$, the
median $1$, and the maximum
$48$. Figure~\ref{fig:num_actions_trajectories} shows
how the number of actions is distributed across the trajectories.

\subsubsection{Relating actions to features}
\label{sec:stat_results}
We now examine the relationships between how much action participants took and
four groups of features---features relating to the trajectory, 
the demographics of the participants, the type of
incident, and the participant's internet browsing behavior---via
four statistical models. 
We consider a
feature to be statistically significant if the significance level ($\alpha$) is less than $0.05$.

Three of the
analyses are over 
the $48$ participants (and $66$ trajectories) who read about an
incident. 
The analysis of
participants' reactions relative to what led them to visit an
incident-related web page is over
52 trajectories, since we removed
trajectories for which we could not determine what led the participant
to visit an incident-related page (i.e., \precursortype was
``Unknown''; see \secref{sec:learning_meth}).

\paragraph{Actions in relation to trajectories}

The first analysis looks at the number of actions participants take
in relation to the following features of the browsing trajectories: 
\precursortype, \homepage, \firstoccurtype, \breachnum, and  \firstoccursentiment.

We modeled this relationship by
a linear regression model, where each trajectory (not each
participant) is one data item.
We use the raw action count as the outcome variable, represented by its Z-score with respect to the mean
number of actions out of all trajectories.
We model each of the three categorical features (with $n$ levels)
with
$n-1$ indicator variables compared to a baseline
level. For \emph{precursor\_type},
we group the five levels into two buckets: ``purposeful,'' for page
visits suggestive of
purposeful browsing (message boards and search results), and
``general,'' for 
 social media, news sites, and general browsing. We also represent the \firstoccursentiment
 feature as the Z-score of the \firstoccursentiment in each trajectory
 with respect to the average sentiment value over all trajectories.

The only factor that was correlated with the number of actions
participants took after visiting an incident-related page was the
sentiment of the content on this first visited page.
A more positive
sentiment was associated with taking more
actions. Specifically, the number of actions increased
by 1.153 standard deviations away from the mean
for every standard deviation increase 
in the value of \firstoccursentiment from its mean ($p=0.02$).
Table~\ref{tab:actions_trajectories} shows the results of the linear regression model.

We conducted a power analysis
to
understand whether our sample was large enougg to reveal effects of a particular size.
Following previous work~\cite{r_pwr, ellis2010essential, al2018effectiveness, tsai2017turtle, mathur2017impact,
nicholson2017can} in aiming for an experimental power of $0.8$, a p-value ($\alpha$) of $0.05$, and a medium effect size, 
we calculated
that a sample size of 34 trajectories
was the minimum to see the desired effects
with our model, a criteria which our sample size of 52 trajectories meets. This suggests
that if our model did not show a factor to be statistically
significant, that factor is likely to not have had a ``medium'' or greater effect.

\begin{table}[]
\centering
\resizebox{\columnwidth}{!}{
\begin{tabular}{|p{45mm}|l|r|r|r|r|r|}
\hline
\textit{\textbf{}} & \textit{\textbf{baseline}\!} & \textit{\textbf{coef.}} & \textit{\textbf{exp(coef.)}} & \textit{\textbf{std.err.}\!} & \textit{\textbf{t}} & \textit{\textbf{p}} \\ \hline
(Intercept) &  & -0.240  & 0.787 & 0.173 &   -1.385 &   0.172 \\ 
\precursortype: purposeful & general & 0.054  &   1.060 & 0.132 &  0.411  & 0.683 \\ 
\homepage: yes & no & -0.025  & 0.975 & 0.124 & -0.206 &  0.838 \\ 
\firstoccurtype: advice & info & -0.117  & 0.890 & 0.172 & -0.678  & 0.501 \\
\firstoccurtype: related & info & -0.167  &  0.846 & 0.137 & -1.226 &  0.226 \\
\breachnum &  & 0.103  &  1.109 & 0.086   & 1.196 & 0.237 \\ 
\textbf{Z(\firstoccursentiment)} &  & \textbf{0.142} & \textbf{1.153} & \textbf{0.060} & \textbf{2.353} & \textbf{0.022} \\ \hline
\end{tabular}}
\vspace*{1mm}
\caption{Linear regression model of the relationship between actions participants took and the trajectories
that led them to finding out about breaches. The outcome variable is log(Z(actions taken) + 1).\label{tab:actions_trajectories}}
\end{table}

\paragraph{Actions in relation to participant demographics}

The second analysis examines the likelihood that participants will
take action (and how many actions they will take) relative 
to participants' demographics, again via a linear
regression model. The demographic features examined are
described in \secref{sec:which_people}.

If a participant had several
trajectories (i.e., visited pages related to more than one breach), we
average their actions across their trajectories.
As before, we model categorical variables as multiple indicator variables. 

No participant descriptors were statistically significant in relation
to the actions participants took.
A power analysis similar to the one we performed for the previous model
showed that the minimum sample size needed
to obtain statistically significant,
medium-sized effects was 36, which our sample of 48 participants exceeded.

\paragraph{Actions in relation to the type of breach}

We examine the relationship between the number of actions taken per trajectory and the type of data 
compromised (\datacomp) using
the Kruskal-Wallis one-way
test of variance~\cite{kruskal1952use}.

The amount of action differed significantly
between categories of the data compromised ($\chi^2 = 19.843$, $df = 3$, $p < 0.001$).
To understand which groups were statistically different from each other
and in what direction, we conducted a post-hoc analysis with pairwise comparisons
using Dunn's test between each group, applying the Bonferroni correction
for each comparison~\cite{dunn1961multiple, bland1995multiple}.

\begin{figure}[tbp]
\centering
\includegraphics[width=\linewidth, height=2in]{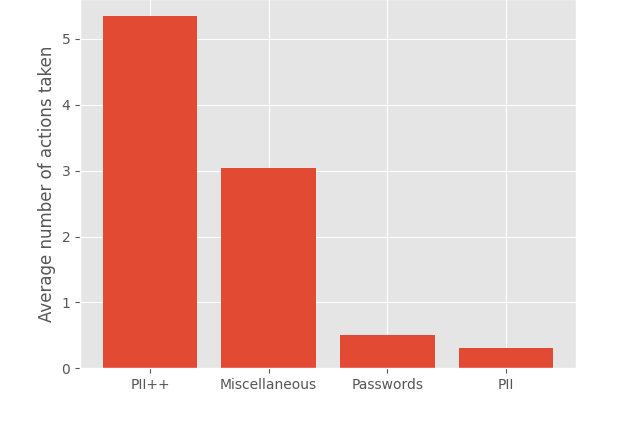}
\caption{Number of actions taken on average for each incident type. \label{fig:all_breaches_actions}}
\vspace{-4mm}
\end{figure}

Participants took an average of 5.35 actions after a PII++-compromised incident,
3.04 after a miscellaneous incident, but only 0.5 and 0.3 after
reading about a passwords or PII incident, respectively.
The greater number of actions taken for PII++ was statistically
significantly higher than for 
passwords ($Z = 3.002, p = 0.02$) or just PII ($Z = 3.85, p < 0.001$).
Figure~\ref{fig:all_breaches_actions} shows a ranking of the average number of actions taken
for the trajectories of each category.

\paragraph{Actions in relation to participants' browsing
  behavior}

Finally, we tested for relationships between the number
of actions people took
and the types
of pages they visited on the web and the amount of technical browsing
(as described in \secref{sec:which_people}).

The linear regression model did not reveal any statistically significant
relationships, although, as before, a power analysis showed that we
had sufficient power to see medium-sized or larger effects.

\subsection{Summary of findings}
In summary, participants came across pages about security incidents through a variety of media in similar proportions.
Most of the times participants came across the page about the incident after browsing deeper through a website,
suggesting that such pages about incidents are not easily accessible (e.g., from a homepage).
Participants were likely to read more about the incident or take an action when the page exhibited a positive sentiment but
no other features were correlated with taking action, implying that the lack of action is nearly universal across our dataset.

\section{Confirming dataset validity}
\label{sec:validity}

Our findings (\secrefs{sec:exp_or_not}{sec:learning_actions}) are based on the browsing
activity collected from one home computer of each
participant. However, participants could have read about incidents or
taken action on other devices, data about which is
not captured in our dataset.

To shed light on how representative our dataset is of 
participants' overall browsing behavior, we collected additional
self-reported data to supplement our main dataset. We conducted a survey of 109 SBO participants who were active
in May 2019, in which we asked about their familiarity
with, and any reactions in response to, several security incidents, as well as
about how much web browsing they perform on which devices. This study was
approved by our institution's review board and the review board of the
SBO's home institution. The survey took between one and five minutes
and participants were compensated with \$5.
Many participants
in our main dataset were not active when we conducted this follow-up survey and vice versa;
hence, we use this survey as 
\emph{a measure of the self-reported behavior
of SBO participants in general}, rather than focusing on specific
individuals who were in both datasets. 

When asked to report the fraction of browsing they performed on their SBO
computers, participants indicated that these were used, on average,
for 57\% of their web browsing. As the amount of browsing on desktop and laptop computers
may have decreased over time in favor of browsing on mobile
devices~\cite{internet_report}, this 57\% is likely a lower bound;
participants earlier in the study likely performed a higher fraction
of their overall browsing on their SBO computers. 

We also asked participants
how often (on a 5-point Likert scale~\cite{vagias2006likert}) they read about security incidents on (1) their SBO computer
and (2) on any other devices. We found no statistically significant
difference between the two distributions
(Kolmogorov-Smirnov~\cite{kolmogorov1933sulla}: $D = 0.056, p = 0.997$). 
We also examined whether the distribution of
browsing on SBO computers vs.\ other devices varied by participant age,
but found no statistically significant indication that it did
(Spearman's correlation test: $S = 180990, p = 0.155$). 
Finally, to gauge the accuracy of the self-reported data, we asked
participants how familiar they were (on a 5-point Likert scale) with five security incidents,
four non-incident-related events, and one fictitious security incident (an Airbnb
social security number breach).
8\% of respondents indicated moderate or extreme familiarity with the
fake Airbnb breach,
suggesting that the self-reported results may slightly exaggerate the
participants' actual familiarity with incidents.

Our results from \secref{sec:exp_or_not} indicate that 25\% of the participants in our main
dataset who were likely to have an Equifax credit report (and therefore, likely to have been affected) 
read about 
the Equifax breach, a surprisingly small percentage.
If we assume that this percentage is computed based on 57\% of 
all browsing, then the actual
percentage of people who read about the breach---if our data included 100\% of all browsing---could
be as high as 36\%, which is still low considering the significance of
the breach.
When asked what action they took following the Equifax breach,
41 participants (38\%) responded that they read about
the breach online and/or visited the Equifax website, with the majority of the rest answering
``didn't do much/didn't do anything''. Five of the 41 respondents
additionally replied that they ``can't remember''
and/or ``didn't do much/didn't do anything'', implying that
the actual number of participants who read about the incident online or visited the website
might be even lower than reported.

Similarly, our results from \secref{sec:exp_or_not} indicate that 2\%
of participants in our main dataset who had a compromised Yahoo!\ password~\cite{yahoo_date_2} read about the Yahoo!\
breach online. Self-reported data also suggests very low awareness: when
asked about reading and reacting to the Yahoo!\ breach, only nine (8\%)
participants answered that they read about the breach online. Two respondents answered
``can't remember'' and/or ``didn't do much/didn't do anything'' in addition to reading online, again
indicating that a lower number of participants than self-reported may have actually read about the incident.

Overall, our results suggest that the browsing data that we use for our
analyses (\secrefs{sec:exp_or_not}{sec:learning_actions}) covers the
majority (with a lower bound of approximately 57\%) of the browsing performed by our participants. While the additional
browsing participants performed on non-SBO devices may dilute some of our
findings about how often people read about incidents, the self-reported data
 supports the big picture: a surprisingly small subset of users
reads about incidents and takes action or tries to learn more about the incidents.

\section{Limitations}
\label{sec:limitations}

Although our work provides valuable insights
into how, and the extent to which, people attempt to make themselves secure after an incident,
it is subject to a few limitations, including those due to the nature of the data collection.

Our dataset contains data about (relatively) few participants due to the difficulty of recruiting participants to the SBO.
However, the SBO data is a tradeoff: it offers rich browsing and password data that is typically infeasible to obtain, at the cost of a limited participant pool and concerns about generalizability. We believe it's the big picture that our results reveal that matters -- that a very small fraction of people seem to read about or try to learn more about security incidents -- rather than the specific percentages involved. Similarly, the browsing history 
may not be representative of all the browsing users do. 
Hence the confirmatory study, which suggests that the high-level results of the original SBO analysis hold: participants were rarely familiar with or read about major security incidents regardless of the devices on which they browsed the internet.

Since browsing history was represented via URLs and page titles, we
could not include analyses that depended on the content of pages that
are dynamic (e.g., social network pages). We also could not
distinguish between content that participants actually
consumed and content they loaded but did not read.
Finally, since participants might open multiple pages in parallel
and click on links in pages opened much earlier, we are
limited in our ability to accurately determine the precursors to the
first incident-related pages participants visited; although in practice
we found only a small number of instances where this was a problem.

The data we analyzed is collected only from Windows computer users.
Users of non-Windows operating systems might 
exhibit behaviors different from the behaviors of the participants in
our dataset. However, as Windows is the dominant OS for personal
computers~\cite{bott_2013}, we do not believe this is likely to fundamentally
affect our findings.

Although data from SBO participants has been used for several security- or
privacy-related studies~\cite{DBLP:conf/ccs/PearmanTNHBCCEF17,
habib2018away, canfield2017replication, forget2016or},
the SBO participants
may be biased towards less privacy- and security-aware
people, given the nature of the SBO data collection
infrastructure. 

Finally, the subsets of participants we used in various specific
analyses were of sufficient size to make uncovering medium-sized
effects likely, but not so large as to reliably discover small-sized
effects. 


\section{Discussion}
\label{sec:discussion}

Our dataset allows a comprehensive view of the browsing behaviors of 303 participants across 44 months. 
Whereas much previous work that investigates people's awareness of and behavior in response to security
breaches and incidents relies on participants' recollections or
reactions to hypothetical scenarios (e.g.,~\cite{mccarney2007hawthorne, forget2016or, habib2018away})---and so can be prone to the
observer effect, among others---our dataset permits investigation of
people's actual awareness of and interactions with incident-related content on the web.

Our findings show that for the people in our dataset, 
the consumption of security and privacy 
incident information is not as prevalent
in people's online activities as is ideal. Even when information was presented and consumed,
people often did not attempt to learn more about the incident or show further interest
in reading about it. Further research is needed to
study how this information can be disseminated more widely and studied in the context of a general population.
To elicit and encourage interest, websites should better 
highlight problems, the implications and risks, and suggestions for staying secure or maintaining privacy in light
of the incident~\cite{zou2019beyond, 
winn2009better, zou2019youmight}.

\paragraph{Improving dissemination of security incident information}
We were surprised to
find that only 48 participants had
visited a web page related to any of the six largely publicized
security incidents.

A majority of the links to the incident-related pages people viewed were encountered on news sites, social media, and online message
boards; and about a third of the articles were related to the Equifax breach.

Our results highlight the challenges of increasing awareness of security incidents and how such information
might be disseminated.
Although the Equifax breach affected more than half of US residents over 18, and
hence likely affected the majority of our participants, 
only 15\% of participants (who were in the study before the breach and in the
study for at least three months afterward)
actually visited an 
article with information about the breach. Without adequate awareness of such incidents,
people are unlikely to understand the importance of safe security behavior
or understand that the implications of the incident may be relevant to them
even if they were not directly affected~\cite{awareness-behavior}.
Although these incidents were highly publicized one
might wonder why people are unlikely to read or learn about
them. 

Recent work by Wash et al.~\cite{DBLP:conf/soups/RaderWB12}
examines what kinds of  ``stories'' are more likely to make an impact on the security behaviors of people, 
and which are more likely to be shared.
We found that a number of
the incident-related articles
participants first discovered were stories about the impacts of the incident, and not about the incident itself.
Additionally, we found that articles or stories with a more positive sentiment, e.g.,
constructive advice and information
as opposed to only warning readers about future risks,
is associated with people exhibiting security-enhancing behaviors.
Taking more action was also correlated with reading about an incident in which
sensitive personal information was disclosed (as with SSNs and
financial information in the Equifax breach), in line
with previous findings about taking action when people perceive
a tangible benefit~\cite{data-breaches}.

Thus, news organizations reporting on security incidents could benefit from research 
on what kinds of stories and content are most likely to influence security behaviors and the further sharing of such content.
Furthermore, while it is expected that people will be more engaged when they have more at stake, challenges remain 
with improving their engagement in the context of everyday services~\cite{shillair2015online}.

\paragraph{Demographic factors related to dissemination and action}

Most of the demographic factors that we examined were not
significantly associated with the likelihood to come across
incident-related articles or with the number of remedial
actions. However,
more technology savvy
participants were much more
likely
to have read about incidents.
This could be because information about incidents is
disseminated more towards technical
audiences, 
perhaps because of the challenges of disseminating
incident information (which may be seen as more technical) on
non-technical outlets. For example, prior work found that such
information is disseminated unevenly based on socio-economic
status~\cite{DBLP:conf/ccs/RedmilesKM16}, which could be linked to
technical savviness. Another potential explanation is that 
technologically savvy people are more receptive to
such information, and so it remains an open challenge to convince less
technologically savvy people about the importance of security incidents and
effectively communicate online risks to them. 
Recent work found that
video communication can raise the saliency of risk for people as
compared to text and concluded that risk communication through videos
might be a more effective way to reach such populations compared to
text~\cite{risk_comm}. 
Thus in addition to exploring what types of ``stories'' are more
effective, further work is needed to explore the medium of
delivering such stories for different populations. For example, research could
investigate how different demographic groups interact with different information dissemination 
channels or properties of shared content.


\section{Conclusion}
\label{sec:conclusion}

Using the actual browsing histories
of 303 participants over four years, we measured how
often people read web pages about security incidents, 
what actions they take after reading, 
and what
factors are associated with how likely they are to read about an incident or take action. 
Our findings are bleak: Only a small minority
(16\%) of participants visited an incident-related web page about one
of six large-scale security incidents. Furthermore, few participants who read about an
incident showed further interest in the incident or took some action by
reading more about it. 

Our results highlight the challenges of increasing awareness of
security incidents and of disseminating information about them. Even
when an incident was highly publicized and participants were likely to
have been affected,
few showed engagement or awareness of the incident, e.g., only 14\%
read about the Equifax breach. Without adequate awareness, it is unlikely that people will act
to improve their security. We found the low rate of discovery,
and of constructive 
action after discovery, to be nearly universal across participants.
Participants with an
affinity for technology were more likely to read about incidents; but
demographic and other factors that we explored either had minor (e.g., age)
or no impact.
When reading web pages that
spoke about the incident in a positive and constructive way, participants
were more likely to try and learn more about the incident or take more action;
but no other factors correlated with taking action. Even though our results
are based on a relatively small population, our results
highlight the need for higher awareness or effective dissemination of security incident
information and advice.

Future work should examine how incident communications
are featured in media frequented by more technology-savvy people and
how their appearance in media with a more general audience could be
modified to improve uptake. Future work should also study content-sharing
platforms used by different demographic groups (e.g., social media) and understand
how the spread of news about incidents can be improved by using such platforms.

\section*{Acknowledgements}
This work was supported in part by the Carnegie Mellon University CyLab Security and Privacy Institute. Parts of the dataset we used were created through work supported by the National Security
Agency under Award No.\ H9823018D0008.
We would also like to thank Jeremy Thomas and Sarah Pearman for help with working with the SBO data.

\bibliographystyle{IEEEtranS}
\bibliography{references}

\appendix
\section{Appendix}

\subsection{Characterizing browsing behavior}
\label{app:browsing_behavior}

To identify topics that characterize participants' browsing, 
we apply the Non-negative Matrix Factorization (NMF) topic-modeling
algorithm~\cite{lee1999learning} to participants' browsing histories.
NMF has been used in prior work for
mining browsing behavior
patterns~\cite{shimada2016browsing, ju2014modeling}.

To build a topic model, we created one document per participant, each consisting of the tokens 
of the Alexa Web Information Services (AWIS) categories of the participant's
web-page visits.
For example, the category for \code{google.com} is
\code{Top/Computers/internet/Searching/Search\_Engines/Google}.

For each participant, we tokenized the AWIS categories
of the domain of each page visit and discarded the ``Top'' token to
create a multiset of tokenized categories. 
If a domain appeared multiple times in a participant's browsing history,
the tokenized AWIS categories appeared an equal number of times.
We then computed the
term frequency-inverse document frequency (TF-IDF)
score~\cite{DBLP:books/daglib/0021593} for each token to produce the
document-token matrix to be used as input by the NMF algorithm.

We applied the NMF topic modeling algorithm to this matrix.
We varied the number of topics from two to $10$ and
identified the optimal number of topics by observing when the
most-frequently occurring tokens in a topic were on average most
similar to each other~\cite{stevens2012exploring,
  pennacchiotti2011investigating}.
To determine this similarity,
we computed the average
of the pairwise cosine similarities of the top 20 tokens within each topic, using a
Word2Vec model~\cite{mikolov2013efficient} trained on the same
documents used to train the topic model, and then averaged
these average similarities.
The average within-topic cosine similarity was highest
when the number of topics was two.

We examined the top 20 tokens in each topic
to determine the themes of the topics. The words in one topic seemed to represent
more leisure-oriented browsing (``Social\_Networking'', ``Shopping'',
etc.)
and the other professional-oriented (``Education'', ``Business'', ``E-mail'', etc.).
As mentioned above, we used the weights of topics in the resulting
document-topic matrix computed over each participant's AWIS multiset as two features.

We also experimented with Latent Dirichlet Allocation (LDA)~\cite{blei2003latent} 
but we observed the resulting topic clusters to not be as coherent as
the ones derived with NMF.

\subsection{Characterizing of browsing behavior of technical or technology-related content}
\label{app:browsing_behavior_technical}
To identify how much of a participant's browsing is technology-related,
we again used the NMF topic modeling algorithm
to build a topic model that has two topics: 
1)~technical or technology-related content and 2)~all other content.

We trained this model using a 1\% sample of each of the participants'
browsing histories and Alexa categories described in \appref{app:browsing_behavior}. 
Here
the input only contains two documents: one for the technical webpages and one for the non-technical
pages. We built each document to be a representation of the content
that is categorized as technical or non-technical by its domain's AWIS category,
i.e., the technical document contains content about web pages that have the 
word ``Technology''
or ``Technical'' in its AWIS category and the non-technical document contains all other content.
We downloaded the content of each web page in the sample with the 
newspaper library~\cite{newspaper},
tokenized each page's content,
and concatenated the tokens from all technical web pages to construct the technical document
and from all other web pages to construct the non-technical document (from a sample of web pages
of the same length as the set of pages in the technical category).
When computing the TF-IDF scores for tokens in each document,
we only include tokens that
appear in one document but not the other.
This way we are able to construct
topics with tokens that are unique to either the technical or non-technical category.

After training the two-topic topic model, we determine the index of the column
in the resulting document-topic matrix that corresponds
to the technical document
and therefore to the technical topic. We apply the trained model on
the sample of each participant's browsing history (a multiset of
tokens of the page content of web pages
with a defined AWIS category). The model computes two weights
for each participant, of which we use the weight
of the technical topic as the feature characterizing the amount of
a participant's browsing related to technical content.

\end{document}